\newcommand{\be}{\begin{equation}}
\newcommand{\ee}{\end{equation}}
\newcommand{\bea}{\begin{eqnarray}}
\newcommand{\eea}{\end{eqnarray}}
\newcommand{\ham}{{\cal H}}

\newcommand{\reff}{{\mbox{eff}}}

\newcommand{\p}{\partial}

\newcommand{\la}{\langle}
\newcommand{\ra}{\rangle}

\newcommand{\nn}{\nonumber \\}

\documentclass{ws-ijmpb}

\begin{document}

\markboth{Sam T. Carr}
{Correlation effects in single-wall carbon nanotubes}

%
\catchline{}{}{}{}{}
%

\title{STRONG CORRELATION EFFECTS IN SINGLE-WALL CARBON NANOTUBES }

\author{SAM T. CARR}

\address{Department of Physics and Astronomy\\ University of Birmingham \\ Birmingham, UK\\
sam.carr@physics.org}

\maketitle

\begin{history}
\received{Day Month Year}
\revised{Day Month Year}
\end{history}

\begin{abstract}
We present a overview of strong correlations in single-wall carbon nanotubes, and an introduction to the techniques used to study them theoretically.  We concentrate on zig-zag nanotubes, although universality dictates that much of the theory can also be applied to armchair or chiral nanotubes.  We show how interaction effects lead to exotic low energy properties and discuss future directions for studies on correlation effects in nanotubes.
\end{abstract}


\section{Introduction}

The world of one spatial dimension has long excited theorists studying many body physics.  The relative importance of interactions as compared to other dimensions means that perturbative approaches often fail, and such models often have strongly correlated ground states not simply related to the single-particle picture of non-interacting electrons.  This is coupled with a wealth of analytic techniques, such as integrability or Bosonisation which generate non-perturbative results for such one dimensional systems, and allow some of the more peculiar properties of strongly correlated systems to be investigated.

Since their discovery\cite{ijima91} in 1991, carbon nanotubes have become an important component of modern physics.  Aside from their obvious possible applications in nanotechnology, they are one of the first example of real one dimensional systems where some of the exciting predictions of strong correlations in one dimension can be seen experimentally.  Indeed, single wall carbon nanotubes (SWNT) are almost ideal clean one dimensional systems, with interaction effects dominating over disorder.

The purpose of this paper is both to give an overview of our current understanding of strong correlation effects in single wall carbon nanotubes and to provide a tutorial introduction to the techniques used to study this for the student who may wish to begin work in this area.  Obviously, the choice of topics included is highly selective in nature and focuses mainly on recent breakthroughs in our understanding, for a much more general and complete introduction to carbon nanotubes, we refer the reader to one of the many existing books or reviews.\cite{sdd98,cbr07,dekker99}

In the past decade or so, there have been many papers dealing with correlation effects in the ground state of models of single-wall carbon nanotubes.\cite{eg97,kbf97,bf97,kll97,lin98,eg98,yo99,oy99,nt03,cgn07}  It was realized early on that the low energy physics of carbon nanotubes is described by ladder models which are well studied in the literature.  However, an isolated nanotube has a Coulomb interaction which is only very weakly screened - thus adding an extra ingredient to existing theory.  Much of the early work was based on renomalization group (RG) analysis, but it was Nersesyan and Tsvelik\cite{nt03} in 2003 who first elucidated the underlying strong coupling phases at half filling in armchair nanotubes by means of a decoupling based on the adiabatic approximation and refermionization of the remaining sectors.  This work was later extended to zigzag nanotubes\cite{cgn07} where the presence of a staggered interaction term competes with the umklapp processes to allow the possibility of a non-trivial quantum critical point.  In the present paper, we review this theory, building it up gradually from all the individual processes present in the Hamiltonian until we put them together to get a full picture of the ground state of single wall carbon nanotubes.

The structure of the paper is as follows: in Section 2 we look at the single-particle picture of electrons hopping on the lattice of a carbon nanotube.  In section 3 we then discuss the important interactions giving rise to strong non-Fermi liquid effects.  In section 4 we then briefly look at some of the experimental evidence for such a theory.  We finish with some conclusions and some appendices detailing some of the technical  and not so technical points involved in the paper.

\section{The non-interacting picture}

\begin{figure}
\begin{center}
\includegraphics[width=1.5in]{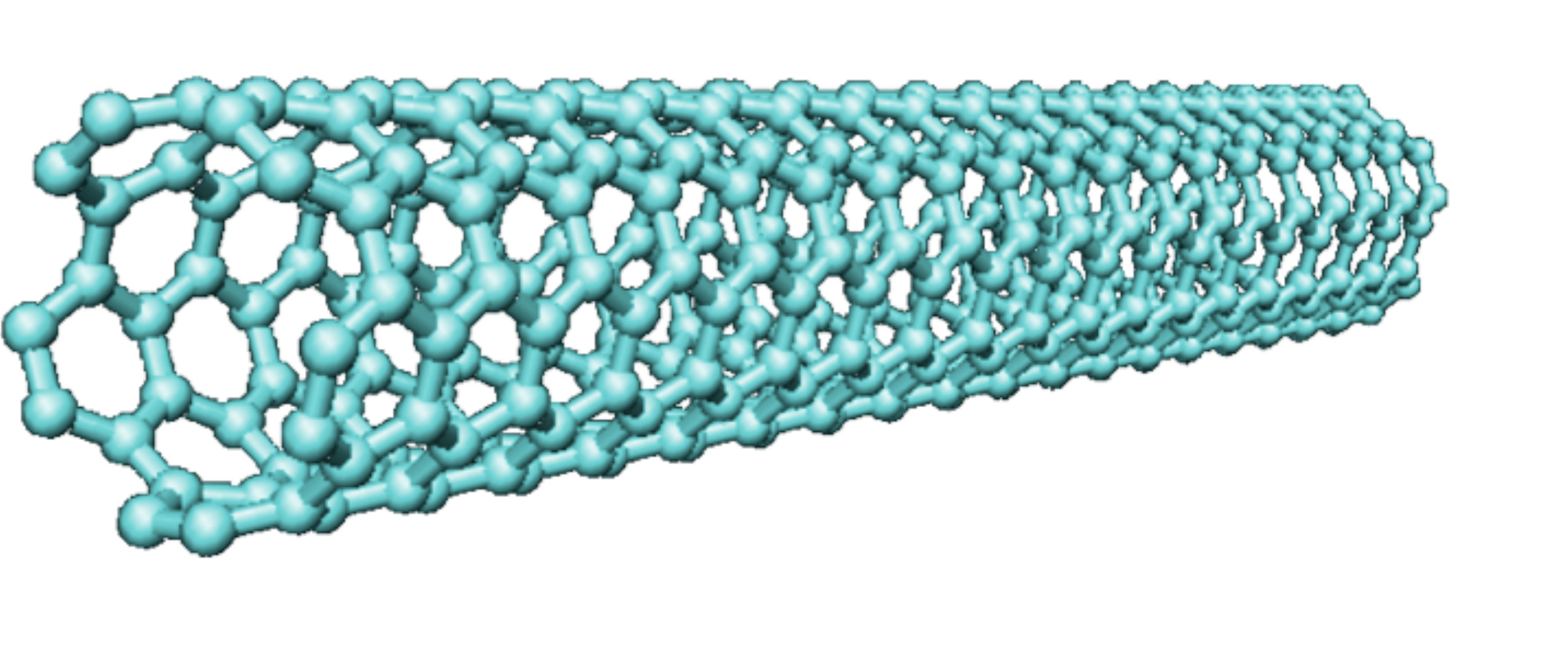}
\hspace{.1in}
\includegraphics[width=1.8in]{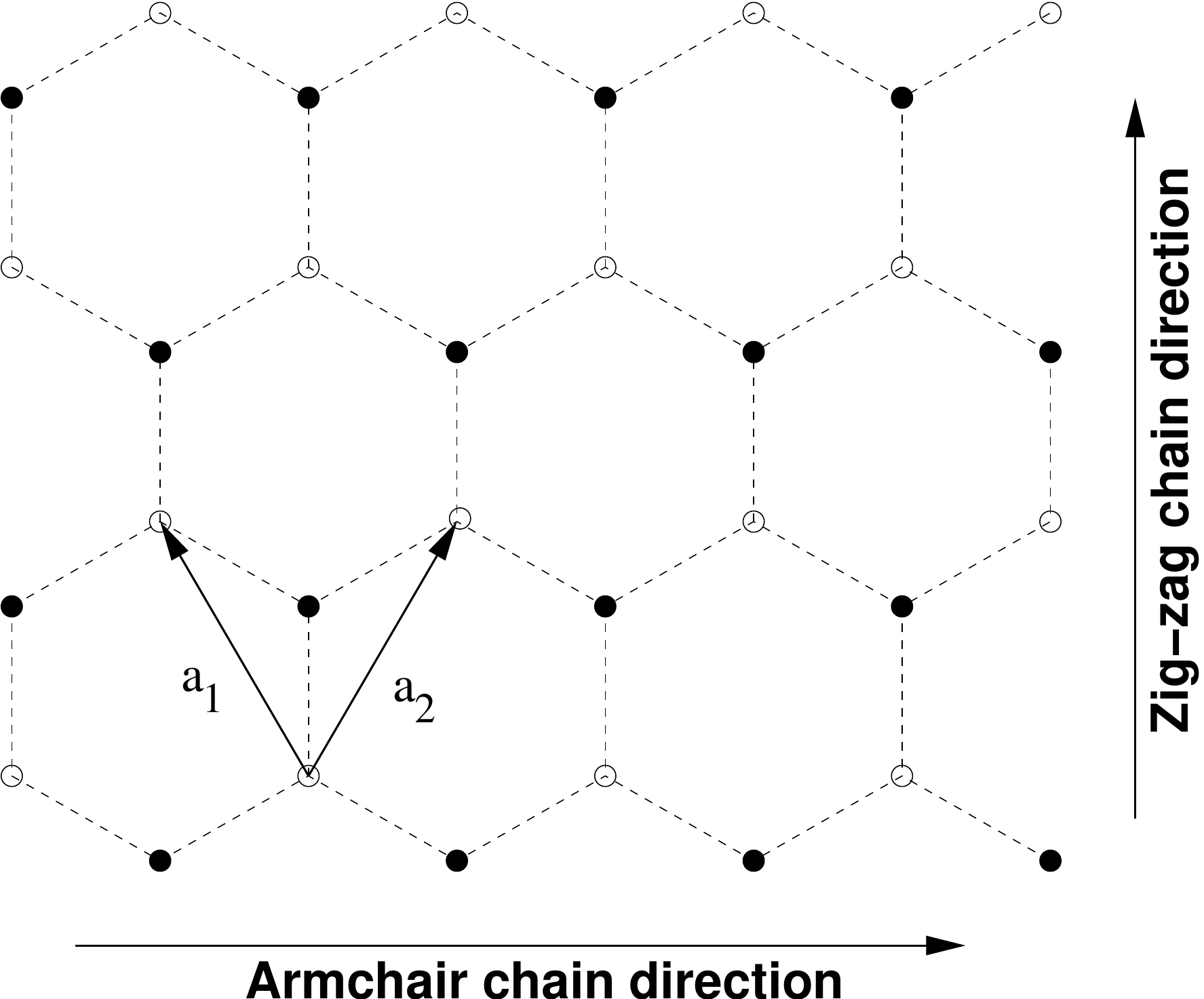}
\hspace{.1in}
\includegraphics[width=0.7in]{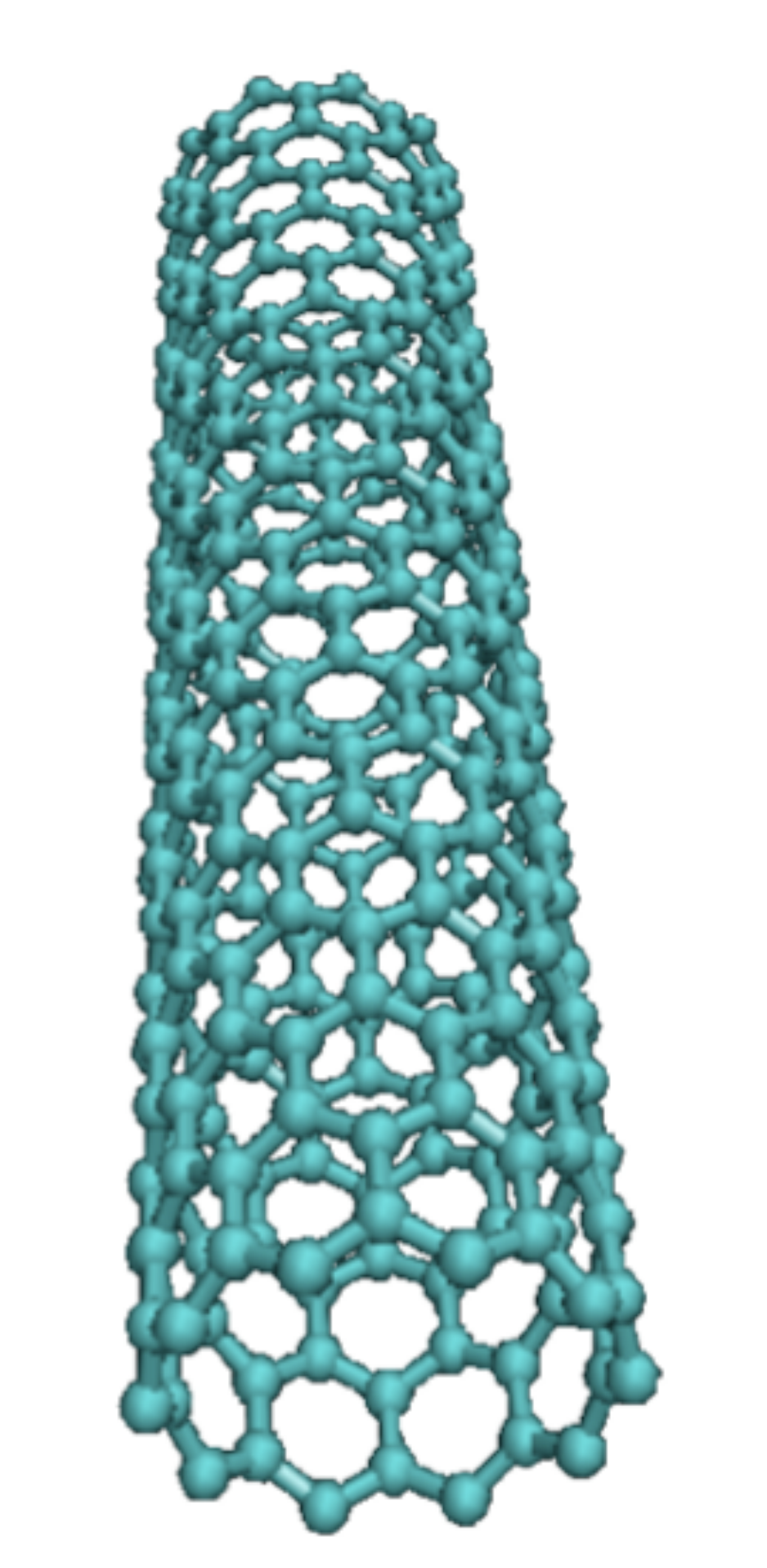}
\caption{The lattice structure of the graphene sheet and armchair (left) and zigzag (right) nanotubes.  The underlying lattice is triangular, with a basis of two carbon atoms per unit cell (open and closed circles).  Note that the two carbon atoms lie at different symmetry positions, but are otherwise identical.  An ($n,m)$ nanotube is formed by wrapping the graphene sheet along the superlattice vector $n\vec{a}_1+m\vec{a}_2$.  There are two special cases with high symmetry: armchair nanotubes when $n=m$, which in the picture are formed by wrapping the bottom of the sheet to the top; and the zigzag nanotubes when $n=-m$, which are formed by wrapping the sheet left to right. }
\end{center}
\end{figure}

Single wall carbon nanotubes are formed by rolling a single layer of graphene into a cylinder, along the direction of the superlattice vector $(n,m)$.  This wrapping vector can be arbitrary, but for the purposes of this review we concentrate on two directions with high symmetry - $(n,n)$ {\em armchair} nanotubes, and $(n,-n)$ {\em zigzag} nanotubes.  In fact, many of the results are universal\cite{oy99} and so can be applied to other {\em chiral} nanotubes as well.

\begin{figure}
\begin{center}
\includegraphics[width=2.5in]{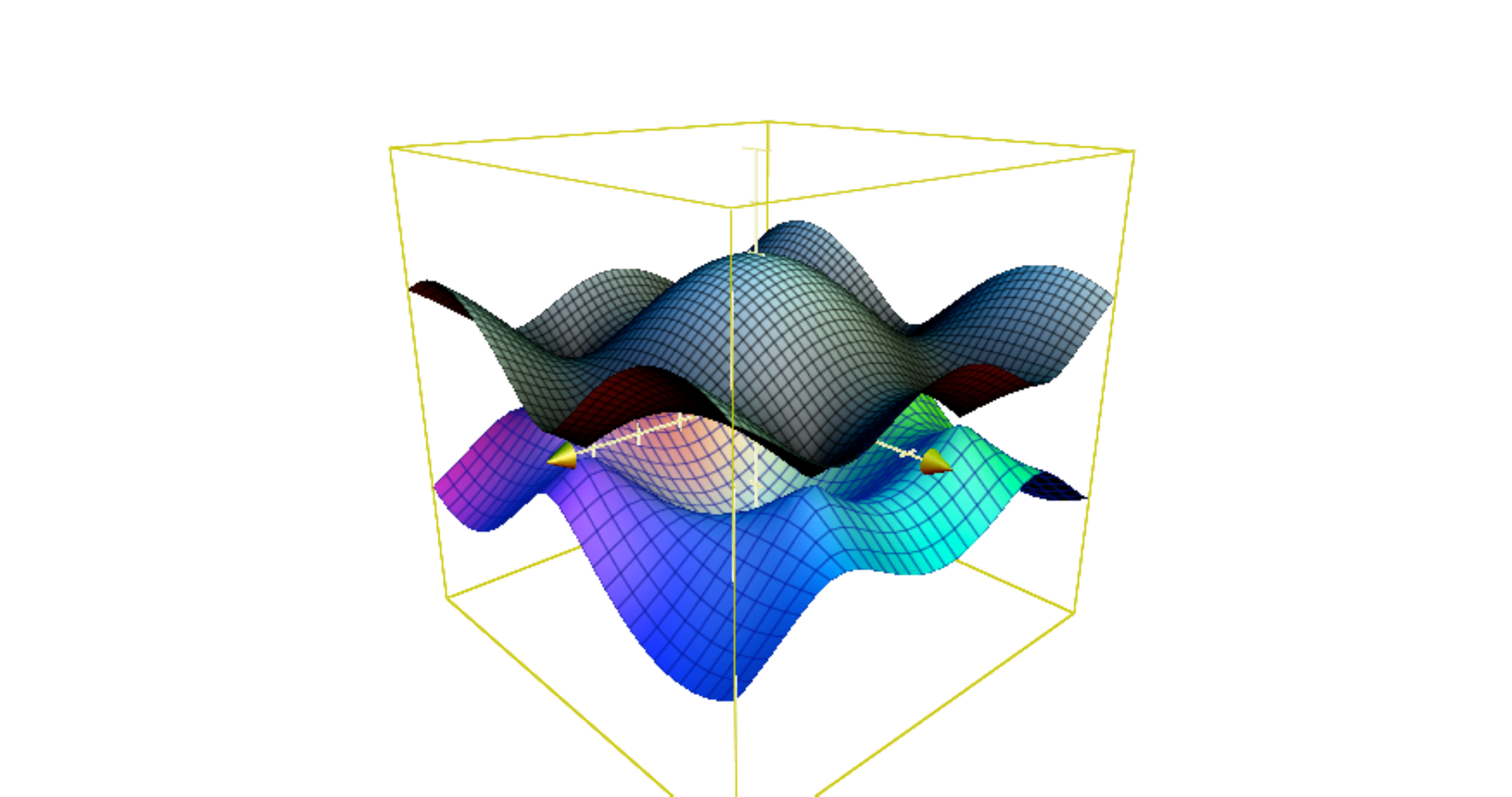}
\caption{Tight binding bandstructure of a single layer of graphene.  At half filling, the Fermi surface is the discrete set of $K$ points, of which only $2$ are independent.   }\label{fig:bandstructure}
\end{center}
\end{figure}

Before looking at nanotubes, one should first understand the band-structure of the underlying two dimensional hexagonal graphene lattice.\cite{wallace}  We start with the tight binding model for the two sublattices, $a$ and $b$
\be
\ham = -t \sum_{\la \vec{i},\vec{j} \ra} c_a^\dagger(\vec{i}) c_b(\vec{j}) + H.c.,\label{eq:originalH}
\ee
where the sum $\la \vec{i},\vec{j}\ra$ is between nearest neighbors on the hexagonal lattice, $c_{a}^\dagger(\vec{i}) [ c_b^\dagger(\vec{i}) ]$ is the creation operator for an electron on sublattice $a$[$b$] at lattice site $\vec{i}$.  A Fourier transform gives
\be
\ham = \left( c_a^\dagger (\vec{k}), c_b^\dagger(\vec{k}) \right) \left( \begin{array}{cc}
0 & 2\cos(\sqrt{3}k_x/2) e^{-ik_y/2} + e^{ik_y}  \\
2\cos(\sqrt{3}k_x/2) e^{ik_y/2} + e^{-ik_y} & 0
\end{array}
\right) \left( \begin{array}{c}
c_a(\vec{k}) \\ c_b(\vec{k} ) \end{array} \right)
\ee
which on diagonalization leads to the spectrum
\be
\epsilon(k_x,k_y)=\pm \sqrt{1+4\cos^2( \sqrt{3} k_x/2) + 4\cos(\sqrt{3} k_x/2) \cos( k_y/2)}.
\ee
This is plotted in Fig. \ref{fig:bandstructure}.  It has the interesting property of being zero only at certain points, known as the $K$ points.  This means that undoped graphene which has precisely one electron per carbon atom so is a half-filled band with chemical potential $\mu=0$ is a semi-metal, with a zero-dimensional Fermi-surface, also often referred to as a zero bandgap semiconductor.  This property is due to the symmetry between the $a$ and $b$ sublattices and remains, even when we go beyond the tight-binding approximation.

When the graphene sheet is wrapped into a nanotube, the allowed momenta in the transverse direction become quantized.  The nanotube can then be an insulator or metal depending on whether the allowed momenta include the $K$ point or not.

Much of the early work on nanotubes involved approximating the low-energy part of the graphene spectra by Dirac-cones, then quantizing the transverse momenta to get the allowed energy bands.  This approach is very useful for showing the universality of conducting nanotubes,\cite{oy99} however in this review we will take a different approach introduced by Lin,\cite{lin98} which shows very transparently the relationship between metallic carbon nanotubes and ladder models.

\subsection{Mapping to ladder models}

\begin{figure}
\begin{center}
\includegraphics[width=1.8in]{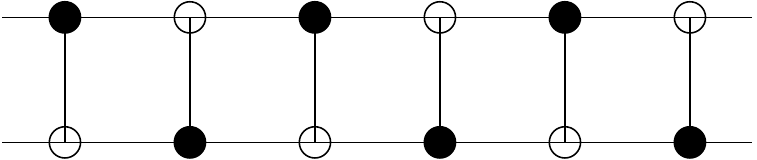}
\hspace{.5in}
\includegraphics[width=1.8in]{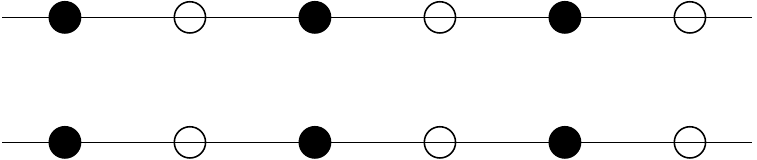} \\
(a)\includegraphics[width=1.2in]{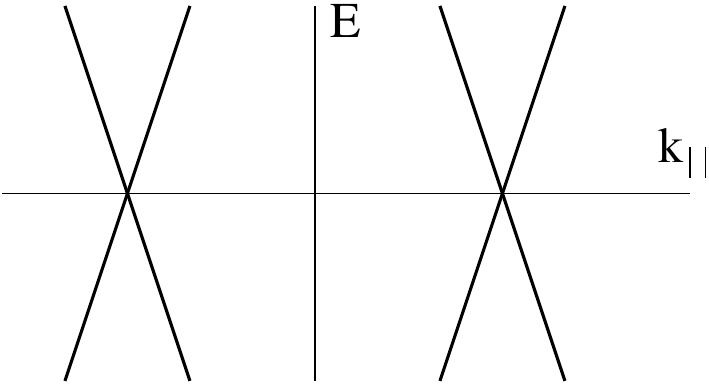}
\hspace{.5in}
(b)\includegraphics[width=1.2in]{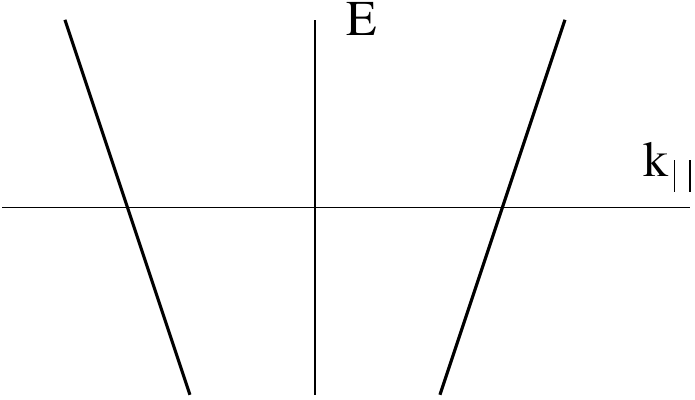} 
\caption{Top - The underlying ladder models for the low-energy properties of (a) armchair and (b) zigzag nanotubes. Bottom - The low energy spectrum of the underlying ladder models.}\label{fig:ladder}
\end{center}
\end{figure}

For the armchair nanotube $(n,n)$, one can make a partial Fourier transform in the transverse direction $y$,
\be
c_{\{a,b\},q}(m_x) = \frac{1}{\sqrt{n}} \sum_{m_y} e^{iq y(m_y)} c_{\{a,b\}} (m_x,m_y).
\ee
The tight binding Hamiltonian Eq. \ref{eq:originalH} then becomes
\be
\ham = -t \sum_{m_x,q} ( e^{iq/2} c_{a,q}^\dagger(m_x) c_{b,q} (m_x+1) + e^{-iq/2} c_{b,q}^\dagger(m_x) c_{a,q} (m_x+1) + e^{-iq} c_{a,q}^\dagger (m_x) c_{b,q}(m_x).
\ee
We know that the only case when we get gapless modes is when the transverse momentum $q=0$, which gives the geometry of the ladder shown in Fig.\ref{fig:ladder}(a).


For the zigzag nanotube, we make an analogous calculation as for the armchair nanotube.  We make a Fourier transform in the transverse direction, which in this case is $x$.  In this case, there is no direct hopping between different states with the same $m_y$, which gives
\be
\ham = -\sum_q \sum_{m_y} t_q(m_y) [ c_q^\dagger (m_y) c_q (m_y+1) + H.c. ]
\ee
where
\be
t_q(m_y) = t_0(q) + (-1)^{m_y} \Delta(q)
\ee
with
\bea
t_0(q) = \frac{1}{2} \left[ t_\| + 2t_\perp \cos (\pi q/n) \right] \\ \nonumber
\Delta(q) = \frac{1}{2} \left[ t_\| - 2t_\perp \cos(\pi q/n) \right]\label{eq:delta}
\eea
and $q$ takes integer values $|q|<n/2$.
In the special case where $n$ is divisible by 3, the two bands $q=\pm Q = \pm n/3$, have a gap $\Delta$ given only by the difference between $t_\|$ and $t_\perp$, i.e. curvature effects, which are small and will be discussed later.  The low energy picture is then one of two chains, illustrated in Fig.\ref{fig:ladder}(b).

\subsection{Low energy spectrum and chiral decomposition}

In a one-dimensional system, the Fermi surface is not a surface but a discrete set of points.  All low-energy physics occurs in the vicinity of these Fermi points.  The case of the zig-zag nanotube is easiest as the low energy theory consists of two decoupled chains.  Each of these chains has the usual tight binding spectrum, and in the vicinity of the Fermi points this can be linearized so that the spectrum looks like two straight lines; one with particles moving to the right at the Fermi velocity $v_F$ and the other with particles moving to the left at this velocity.

It is convenient to expand the original lattice operators in terms of smooth fields living at one of these two Fermi points - a technique known as chiral decomposition:
\be
c_{\mu\sigma}(l) \rightarrow \sqrt{b} \left[
e^{i k_F b l } R_{\mu\sigma}(x) + e^{-i k_F b l } L_{\mu\sigma}(x) \right], \;\;\; x=bl,
\ee
where $\mu$ is the band-index indicating which 'leg' of the ladder we are sitting on, $\sigma$ is the spin index, and $b$ is the effective lattice spacing.  Substituting this into the tight-binding Hamiltonian for the ladder model gives the effective Hamiltonian $H_0 = \int\, dx \ham_0$, where
\be
\ham_0 = -iv_F \sum_{\mu\sigma} \left( R_{\mu\sigma}^\dagger \p_x R_{\mu\sigma} -L_{\mu\sigma}^\dagger \p_x L_{\mu\sigma} \right).
\ee

For the armchair nanotubes, the fact that the ladder is two strongly coupled chains means that before we perform the chiral decomposition, we must first diagonalise the Hamiltonian.  This is easily done by taking the bonding and antibonding combinations of the two legs of the ladder
$c_{\pm , \sigma} = \frac{1}{\sqrt{2}} \left( c_{1,\sigma} \pm c_{2,\sigma} \right).$
The spectrum is then shown in Fig.~\ref{fig:ladder}(a), and clearly shows its relationship to the Dirac cones of graphene.

\section{Interactions}

It is well known that reduced dimensions enhance the effects of interactions; in one spatial dimension Fermi-liquid breaks down completely, even for arbitrarily small interactions and a new paradigm is needed: the Luttinger liquid.  We will see how such considerations apply to nanotubes.

It turns out that for undoped isolated nanotubes, there are three important interactions to take into account:
\begin{enumerate}
\item Unscreened long-range Coulomb interaction,
\item Umklapp interactions,
\item Dimerization of interactions due to the way we break the $C_3$ symmetry of the underlying graphene sheet.
\end{enumerate}

We will examine each of these interactions in detail in the following sections, to identify both their origins in the underlying theory, and the effect they have on the low energy electronic properties of single wall carbon nanotubes.  The main theoretical tool used to handle interactions in one-dimensional systems is bosonization (see Appendix \ref{sec:boz}).  However at all stages we supplement the formal arguments with cartoons (see Appendix \ref{sec:cartoon}) to aid the reader less familiar with low-dimensional field theory.

As demonstrated in Appendix \ref{sec:boz},  bosonization is based on the correspondence
\bea
R_{\mu\sigma} (x) = \frac{1}{\sqrt{2\pi\alpha}} \exp \left\{ -i\sqrt{\pi}\left[ \Phi_{\mu\sigma}(x) - \Theta_{\mu\sigma}(x) \right] \right\}, \\ \nonumber
L_{\mu\sigma} (x) = \frac{1}{\sqrt{2\pi\alpha}} \exp \left\{ -i\sqrt{\pi}\left[ \Phi_{\mu\sigma}(x) +\Theta_{\mu\sigma}(x) \right] \right\},
\eea
where $\Phi_{\mu\sigma}(x)$ and $\Theta_{\mu\sigma}(x)$ are a pair of mutually self dual scaler fields with the commutation relation $[\Phi_{\mu\sigma}(x) ,  \Theta_{\mu\sigma}(y)]=i\delta_{\mu\mu'}\delta_{\sigma\sigma'}$ when $y>x$ and zero otherwise, and  $\alpha$ is a short-distance cutoff.  Basically, this describes the excitation spectrum of the one-dimensional Fermi gas as propagating density fluctuations.  These fluctuations can be categorized by using the linear combinations
\be
\Phi_c^\pm = \frac{1}{2} \left( \Phi_{+\uparrow}+ \Phi_{+\downarrow} \pm \Phi_{-\uparrow} \pm \Phi_{-\downarrow} \right), \;\;\;
\Phi_s^\pm = \frac{1}{2} \left( \Phi_{+\uparrow}- \Phi_{+\downarrow} \pm \Phi_{-\uparrow} \mp \Phi_{-\downarrow} \right),
\ee
which describe total $(+)$ and relative $(-)$ charge $(c)$ and spin $(s)$ fluctuations.

The kinetic part of the Hamiltonian then becomes
\be
\ham_{kin} = \frac{v_F}{2} \sum_{a=(c^\pm,s^\pm)} \left[ \left( \p_x \Theta_a \right)^2
+ \left( \p_x \Phi_a \right)^2 \right],
\ee
which is an independent Gaussian model for each of the modes.
We will see later that this decomposition of modes is very useful in the charge sector, however it is somewhat misleading in the spin sector, where the sensible mode decomposition is into a singlet and triplet channel, and not the two independent spin-half opjects.  This is an artifact of using the standard bosonization which is Abelian on a model with a non-Abelian $SU(2)$ spin symmetry - however we will see how to fix this later and make the symmetries of the model more apparent.

\subsection{Long range Coulomb}\label{sec:Coulomb}

In most situations, we are used to dealing with short range interactions as the long-range tail of the Coulomb interaction is rapidly screened.  In a nanotube, some level of screening can indeed take place through a substrate, however for an isolated nanotube this is weak and one must consider the full long-range nature of the Coulomb interaction.

We begin by looking at what the long-range Coulomb interaction does to the low-energy excitations in carbon nanotubes.  In terms of the bosonic fields, the density $\rho(x)$ at a point $x$ along the nanotube is given by
\be
\rho(x) = \frac{1}{\pi} \p_x \Phi_c^+ (x)
\ee
and depends only on the total charge sector of the theory, with there being no difference between armchair and zigzag nanotubes. The Coulomb interaction is therefore
\be
H_{Coul} = \frac{2e^2}{\pi} \int \, dx \int \, dy \frac{\p_x \Phi_c^+(x) \p_y \Phi_c^+ (y)}{|x-y|}.
\ee
Now, the Coulomb interaction is eventually screened we make the reasonable assumption that the screening length $R_s$ is long compared with the radius of nanotube $R$, but short compared with it's overall length $L$, i.e. $R \ll R_s \ll L$.  Then, assuming that the electrons do not form a Wigner crystal, the underlying density is uniform along the nanotube and so the above expression can be approximated by\cite{kbf97}
\be
H_{Coul} \approx \frac{e^2}{\pi^2} \ln (R_s/R) \int dx\,  \left( \p_x \Phi_c^+ \right)^2.
\ee
The sum of the kinetic and Coulomb parts of the Hamiltonian in the total charge sector then remains a Gaussian model, but with the fields rescaled - i.e. it is a Tomonaga-Luttinger liquid
\be
H_\rho^+ = \frac{v_\rho^+}{2} \int dx \,
\left[
\frac{1}{K_c^+} \left( \p_x \Theta_\rho^+ \right)^2 +
K_c^+ \left( \p_x \Phi_\rho^+ \right)^2 \right],\label{eq:Hcoul}
\ee
with velocity
\be
v_\rho^+ = v_F \sqrt{1 + \frac{8e^2}{v_F}\ln\left(\frac{R_s}{R}\right) },
\ee
and Luttinger liquid parameter $K_c^+ = v_F/v_\rho^+$.  The long-range nature of the Coulomb interaction therefore gives a strong renormalization of the field, giving a strong deviation from the non-interacting scaling behaviour of quantities such as tunneling density of states.  A cartoon picture of this state is given in Appendix \ref{sec:plasmons}.

The Luttinger liquid parameter can also be related to mesoscopic properties of the nanotube which experimentally has a finite length.\cite{kbf97}  Firstly, there is the level spacing
\be
\Delta E \sim v_F / L,
\ee
and secondly, there is the (capacitive) charging energy
\be
E_c \sim \frac{e^2}{L} \ln \frac{R_s}{R}.
\ee
This means that the Luttinger liquid parameter may be expressed as
\be
K = \left( 1 + 4 E_c / \Delta E \right)^{-1/2},
\ee
which allows an independent experimental determination of $K$ to compare with the theory.

Such experiments have been effectuated\cite{bock99} by measuring the tunneling conductance.  In a Luttinger liquid the tunneling density of states is given by (see Appendix~\ref{sec:tdos})
\be
\nu(\omega) \sim |\omega - \epsilon_F|^\alpha
\ee
where the exponent $\alpha$ depends on the Luttinger liquid parameter, and also whether the tunneling is into the end of the middle of the tube:
\bea
\alpha_{bulk} = (K^{-1} + K -2)/8 \\ \nonumber
\alpha_{end} = (K^{-1} - 1) /4.
\eea
This power law behaviour for the density of states gives similar power law behaviours for the transport properties - the conductance at small bias voltage $V$ is $G(T) \propto T^\alpha$ and at large bias voltage we have $dI/dV \propto V^\alpha$.  We will remark more fully on the results of the experiments in Section \ref{sec:expt}.

As a final remark about the effects of the unscreened Coulomb interaction, the strong renormalization in the total charge sector also means that the velocity of the plasmons  $v_\rho^+$ is much greater than the bare Fermi velocity $v_F$ of the excitations in the other sectors of the theory. This can be understood by considering that the strong interaction gives a very high 'charge stiffness', i.e. changing the total density anywhere in the nanotube is difficult because of the long range Coulomb interaction, and therefore excitations travel very quickly. This gives rise to a strong separation of energy scales between the total charge sector and all other sectors of the theory, allowing one to make the adiabatic approximation which will become essential in the analysis of the other interactions.

\subsection{Long range Coulomb + umklapp}

The previous section outlined the theory  for metallic nanotubes with an unscreened density-density Coulomb interaction present.  It turns out that such a theory is enough for sufficiently doped nanotubes.  However an undoped nanotube is exactly half-filled and this means one must also consider the physics of Mott insulators - in other words one must think about the more detailed structure of the interaction arising because the underlying model is actually on a lattice.  At half filling, the Fermi wavevector $k_F = \pi/2$, which means that terms involving $4k_F = 2\pi$ are once again uniform (in fact, they are the traditional umklapp terms), and must be included.  For a cartoon representation of this physics, see appendices \ref{sec:mott}, \ref{sec:soliton} and \ref{sec:confinement}.

We will work with zigzag nanotubes, although the theory as it stands in this section is actually more relevant for the armchair case.  The reason for this is twofold: firstly it is easier to work with the zigzag nanotubes as the starting point is the ladder model with no direct tunneling between the two chains, in contrast to the armchair case where the tunneling is stronger than all of the interactions.  In fact, it can be shown\cite{wlf03} that these two limits are identical within a duality transformation so the theory presented here can be equally well applied to the armchair case.  So while the physics of these two theories is identical, the operator correspondence to the original wrapped graphene sheet is not - so we present the results in the form that will be useful in later sections when we dig deeper into the physics of zigzag nanotubes.  For an analogous calculation to this section performed directly on armchair nanotubes, see Nersesyan and Tsvelik.\cite{nt03}

One can construct this theory in a number of ways, for example, by applying the full structure of the Coulomb interaction on a lattice and proceeding from there, or by constructing all allowed terms in the effective Hamiltonian.  Here, we will use a variation of the latter method\cite{nt03,cgn07} by constructing the minimal model on the original hexagonal lattice which does not contain any spurious symmetries, and then using the mapping developed above, compute the low energy effective Hamiltonian.

It turns out that the minimal model contains both an onsite interaction, $U$, and a nearest neighbour interaction, $V$.  By considering these interactions, along with the long range tail of the unscreened Coulomb interaction discussed previously, we can construct a complete theory of the low energy physics of armchair single wall carbon nanotubes.

\begin{figure}
\begin{center}
\includegraphics[width=3in]{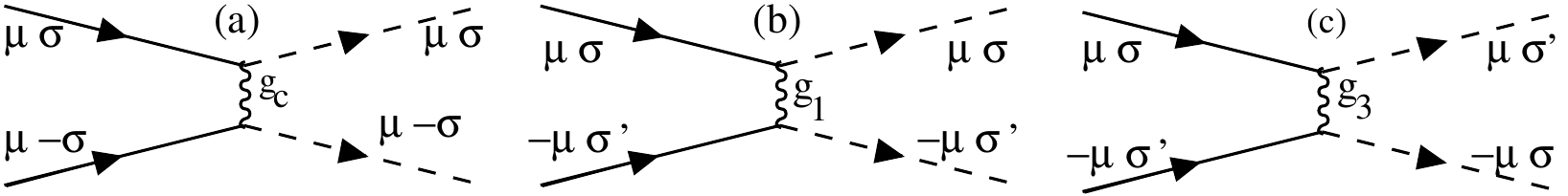}
\caption{The umklapp interaction terms.  Solid lines represent right moving particles, dashed lines left moving particles.  $\mu$ is the chain index, and $\sigma$ the spin index.}\label{fig:interactions}
\end{center}
\end{figure}

Mapping the $U-V$ model on the original graphene sheet onto the ladder with two chains $\mu=\pm$, the interaction is
\bea
H_U &=& \frac{U}{n} \sum_{i\mu} n_{\mu\uparrow}(i) n_{\mu\downarrow}(i) + 
\frac{U}{n} \sum_{i\mu} c_{\mu\uparrow}^\dagger c_{-\mu\uparrow} c_{-\mu\downarrow}^\dagger c_{\mu\downarrow} \nn
H_V &=& \frac{3V}{2n} \sum_i n(i) n(i+1)\label{eq:int_smooth}
\eea
where $n_{\mu\sigma}=c_{\mu\sigma}^\dagger c_{\mu\sigma}$ and $n=\sum_{\mu\sigma} n_{\mu\sigma}$.  By carrying out first the chiral decomposition on these terms, one obtains a jungle of terms.  However, it turns out that because of the long range Coulomb interaction, which makes $K_c^+$ very small, the most important terms are the umklapp terms - represented pictorially in Fig. \ref{fig:interactions}.  This leads to a bosonized Hamiltonian:
\bea
\ham &=& \ham_{kin} + \ham_{Coul} + \ham_{umk} \\ \nonumber
\ham_{umk} &=& -\frac{1}{2 (\pi \alpha)^2} \cos\left(\sqrt{4\pi} \Phi_c^+ \right) \left[
g_c \cos\left(\sqrt{4\pi}\Phi_c^-\right)+
 \right. \\ \nonumber
&& \left.
 (g_3-g_1) \cos\left(\sqrt{4\pi}\Phi_s^+\right)
+ g_3  \cos\left(\sqrt{4\pi}\Theta_s^-\right) - g_1  \cos\left(\sqrt{4\pi}\Phi_s^-\right) \right] + \ldots
\label{eq:fullham}
\eea
The scaling dimension of each of these terms is $1+K_c^+$, and as $K_c^+$ is rather small from the unscreened Coulomb interaction, they are strongly relevant.  The remaining terms not written here which are a combination of various backscattering terms all have scaling dimension $\approx 2$, so are marginal, and will not have any effect on the physics.

Within the $U-V$ model, $g_c = g_1 = (U-3V)/n$ and $g_3 = U/n$, however we will leave $g_c$ and $g_1$ separate for completeness.  The effect of this term was first investigated via a renormalization group analysis,\cite{yo99,eg97} however we will take a different approach\cite{nt03} which gives far more insight into the nature of the possible strong coupling ground states of this model.  This approach which centers around an adiabatic approximation is based on the strong renormalization of the $\Phi_c^+$ field, and the fact that the excitations of this field travel with a velocity much greater than the bare Fermi velocity $v_c^+ \gg v_F$.  This means that from the point of view of the 'fast' field $\Phi_c^+$, all of the other fields look static, so can be replaced by their instantaneous expectation values.  The effective problem in the total charge field is therefore a sine-Gordon model (see Appendix \ref{sec:sG})
\be
\ham [ \Phi_c^+] = \frac{v_c^+}{2} \left[ (\Pi_c^+)^2 + (\p_x \Phi_c^+)^2 \right] + \frac{\la U \ra}{2\pi\alpha} \cos \beta\phi
\ee
where
\be
U = g_c \cos\left(\sqrt{4\pi}\Phi_c^-\right)+
 (g_3-g_1) \cos\left(\sqrt{4\pi}\Phi_s^+\right)
+ g_3  \cos\left(\sqrt{4\pi}\Theta_s^-\right) - g_1  \cos\left(\sqrt{4\pi}\Phi_s^-\right) 
\ee
and $\beta=\sqrt{4\pi K_c^+}$.  The small-$\beta$ sine-Gordon model has a large gap to solitonic excitations of the $\Phi_c^+$ field, but the lowest energy excitations are the breather modes  Translated into physical terms they are the plasmon's of the original Luttinger liquid, but are now gapped.  We will come back to the structure of the interaction (\ref{eq:fullham}) to study the full excitation spectrum of the carbon nanotubes but for now we will concentrate on the collective excitations with energies below the gapped plasmons.  In this region, we can integrate out the $\Phi_c^+$ field, which in practice means we can replace $\cos\sqrt{4\pi}\Phi_c^+$ by its expectation value - note that this approximation is also backed up by the RG analysis.\cite{yo99}  This decouples the remaining modes, which is clearest to see after refermionization (see Appendix \ref{sec:refermionization})
\bea
\ham_{eff} &=& 
i v_F \left[ - R_f^\dagger \p_x R_f + L_f^\dagger \p_x L_f \right]
- i m_f (R^\dagger_f L_f - L_f^\dagger R_f ) \nn
&+& \frac{i v_F}{2} \sum_{j=0}^{3} ( - \chi^j_R \p_x \chi^j_r + \chi^j_L \p_x \chi^j_L )
- im_t\sum_{a=1}^3\chi_R^a\chi_L^a -im_s \chi_R^0\chi_L^0 
\eea
where
\bea
m_f &=& \frac{v_F g_c}{\pi \alpha v_c^+}=\frac{v_F (U-3V)}{\pi \alpha v_c^+ n}, \nn
m_t &=& \frac{v_F (g_3-g_1)}{\pi \alpha v_c^+} = \frac{3v_F V }{\pi \alpha v_c^+ n}, \nn
m_s &=& \frac{v_F ( -g_3-g_1)}{\pi \alpha v_c^+} = \frac{-v_F (2U-3V)}{\pi \alpha v_c^+ n}
\eea
The Dirac fermions $R_f$ and $L_f$ come from the refermionization of the field $\Phi_c^-$ which governs {\em vortex} excitations.  The refermionization of the spin sector in terms of four real (Majorana) fermions $\chi^j_{R,L}$, $j=0\ldots 3$ clearly shows the original $SU(2)$ spin symmetry is still present, and the excitations separate into a triplet ($j=1\ldots 3$) and a singlet ($j=0$) excitation.

The operators $\sigma$ and $\mu$ are Ising operators (see Appendix \ref{sec:refermionization}), and we use the correspondence:
\bea
\cos(\sqrt{\pi}\phi_s^+) = \sigma_1\sigma_2, & \;\; &  \cos(\sqrt{\pi}\theta_s^+) = \mu_1\sigma_2 \nn
\sin(\sqrt{\pi}\phi_s^+) = \mu_1\mu_2, & \;\;&  \sin(\sqrt{\pi}\theta_s^+) = \sigma_1\mu_2 \nn
\cos(\sqrt{\pi}\phi_s^-) = \sigma_0\sigma_3,  &\;\;&\cos(\sqrt{\pi}\theta_s^-) = \mu_0\sigma_3 \nn
\sin(\sqrt{\pi}\phi_s^-) = \mu_0\mu_3, & \;\;& \sin(\sqrt{\pi}\theta_s^-) = \sigma_0\mu_3
\eea
In general, one has that $\la \sigma_i \ra \ne 0$ if the corresponding Majorana mass $m_i<0$, and $\la \mu_i \ra \ne 0$ if $m_i>0$.

There are eight possible phases depending on the relative signs of $m_f$, $m_t$ and $m_s$:
\begin{center}
\begin{tabular}{ccccc}\hline
A & $m_f>0, m_s<0, m_t<0$ & $\la \cos(\sqrt{\pi}\phi_c^-)\ra \ne 0$ & $\la \sigma_a \ra\ne 0$ & $\la \sigma_0 \ra\ne 0$ \\ 
B & $m_f>0, m_s>0, m_t<0$& $\la \cos(\sqrt{\pi}\phi_c^-)\ra \ne 0$ & $\la \mu_a \ra\ne 0$ & $\la \sigma_0 \ra\ne 0$ \\
C & $m_f<0, m_s>0, m_t<0$& $\la \sin(\sqrt{\pi}\phi_c^-)\ra \ne 0$ & $\la \mu_a \ra\ne 0$ & $\la \sigma_0 \ra\ne 0$ \\
D & $m_f<0, m_s>0, m_t>0$& $\la \sin(\sqrt{\pi}\phi_c^-)\ra \ne 0$ & $\la \mu_a \ra\ne 0$ & $\la \mu_0 \ra\ne 0$ \\
E & $m_f<0, m_s<0, m_t>0$& $\la \sin(\sqrt{\pi}\phi_c^-)\ra \ne 0$ & $\la \sigma_a \ra\ne 0$ & $\la \mu_0 \ra\ne 0$ \\
F & $m_f>0, m_s<0, m_t>0$& $\la \cos(\sqrt{\pi}\phi_c^-)\ra \ne 0$ & $\la \sigma_a \ra\ne 0$ & $\la \mu_0 \ra\ne 0$ \\
G & $m_f>0, m_s<0, m_t>0$& $\la \cos(\sqrt{\pi}\phi_c^-)\ra \ne 0$ & $\la \sigma_a \ra\ne 0$ & $\la \mu_0 \ra\ne 0$ \\
H & $m_f<0, m_s>0, m_t<0$& $\la \sin(\sqrt{\pi}\phi_c^-)\ra \ne 0$ & $\la \mu_a \ra\ne 0$ & $\la \sigma_0 \ra\ne 0$ \\\hline
\end{tabular}
\end{center}

Possibilities G and H are prohibited if $g_c=g_1$ as would be the case for the $U$-$V$ model of the interactions in nanotubes. These possibilities were all categorized by Nersesyan and Tsvelik\cite{nt03} thus giving the phase diagram (in parameter space) of armchair single wall carbon nanotubes (although beware of the difference in sign conventions between our derivation for zigzag nanotubes, and their derivation for the armchair case).  For an unadulterated nanotube one would expect $U \gg V>0$, i.e. the onsite repulsion is much larger than the off-site repulsion (and both are repulsive), which forces nanotubes to be in our phase F - so we will concentrate on only this phase.

Due to the opposite signs in the masses of the singlet and triplet spin excitation modes, there are no local operators on the nanotube that have a non-zero expectation value in this phase, although there are obviously a choice of non-local order parameters.  This satisfies the definition of a Haldane spin liquid phase i.e. a gapped phase but with no long range order.  The phase can be understood further by defining the staggered magnetization as a suitably averaged difference between the local spin densities on the two sublattices of the original graphene sheet
\be
\vec{n}^-(\vec{r}) = \vec{S}_a(\vec{r}) - \vec{S}_b(\vec{r}).
\ee
Projecting this onto the low-energy sector of the model gives
\be
\vec{n}^- \sim \cos(\sqrt{\pi}\Phi_c^-) \mu_0 \left(\begin{array}{l}
\mu_1\sigma_2\sigma_3\\
\sigma_1\mu_2\sigma_3\\
\sigma_1\sigma_2\mu_3
\end{array} \right)
\sim \la \cos(\sqrt{\pi}\Phi_c^-) \ra \la \mu_0 \ra \la \sigma_a \ra^2 \left(\begin{array}{l}
\mu_1\\ \mu_2 \\ \mu_3 \end{array}\right).
\ee
The fact that this contains only one of the Ising models means that the two-point correlation function of $\vec{n}^-$ will display a coherent magnon peak, with a mass gap of $|m_t|$.

We finally note that as well as the gapped plasmons and the gapped collective modes in other channels, the model itself has quasi-particle excitations, which can be seen by going back to the original Hamiltonian Eq. \ref{eq:fullham}.  A half-soliton simultaneously in each of the modes leaves the Hamiltonian invariant, and these are excitations carrying charge $e$ and spin $1/2$, i.e. they are quasiparticle excitations.  We simply note that these are gapped on a large energy scale - the most important excitations at low energies are the collective spin excitations.

\subsection{Competition of umklapp and single particle gap}\label{sec:zzqpt}

\begin{figure}
\begin{center}
\includegraphics[width=1in]{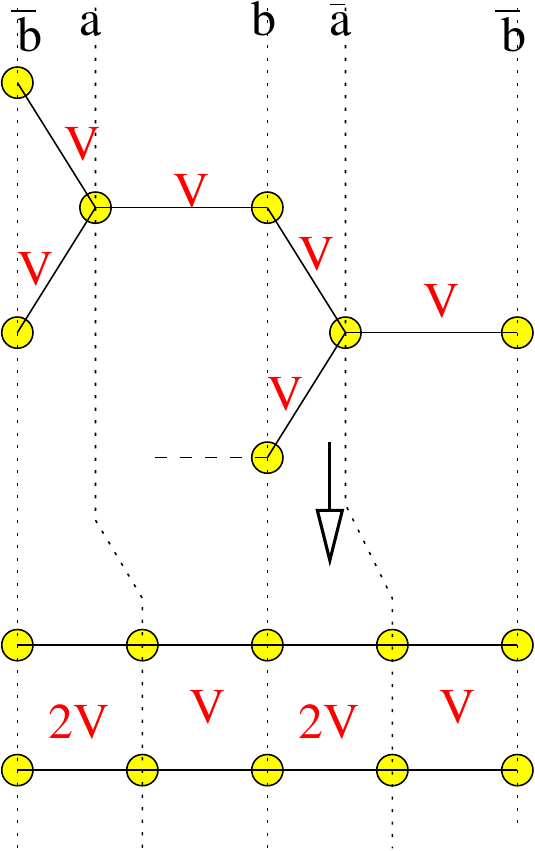}
\caption{The origin of the staggered interaction.  When the graphene sheet is wrapped into a zigzag nanotube, the interactions do not have a reflection symmetry about any of the lattice points.}\label{fig:stag1}
\end{center}
\end{figure}

While the previous section should give the full story for armchair nanotubes, which have no single particle gap, zigzag nanotubes have a curvature induced gap.  As seen in Eq. \ref{eq:delta}, this comes about from a slight difference between $t_\|$ which goes directly along the nanotube and $t_\perp$ which has to partially curve around the waist.  It adds a term in the Hamiltonian
\be
H_{\rm dim} = \Delta \sum_{\mu\sigma} \sum_n
(-1)^n [ c^\dagger_{\mu\sigma}(n) c_{\mu\sigma}(n+1) + H.c ].\label{eq:dim1}
\ee
Now, it was shown by Kane and Mele\cite{km97} that this gap is small  (of order $1/n^2$) while the interaction gaps are all of order $1/n$.  However, there is another process\cite{cgn07} that generates an identical term in the Hamiltonian, but is of order $1/n$.  This term is the staggered interaction, shown schematically in Fig. \ref{fig:stag1}.  In fact, being more careful about mapping the $U-V$ model of the zigzag nanotube onto the low energy sector, one sees that not only the terms in Eq. \ref{eq:int_smooth} are generated, but also the staggered term
\be
H_s = -\frac{V}{2n} \sum_l (-1)^l n(l) n(l+1) + \frac{V}{n} \sum_{l\mu\sigma\sigma'}
(-1)^l c_{\mu\sigma}^\dagger(l) c_{-\mu\sigma}(l) c_{-\mu\sigma'}^\dagger(l+1) c_{\mu\sigma'}(l+1).\label{eq:dimstag}
\ee

This staggered term is from the nearest neighbour interaction $V$, and is entirely due to the way the $C_3$ symmetry of the original hexagonal lattice is broken by wrapping it into a nanotube - turning the figure on its side shows that no such term is present in the armchair case.

Again, going through the chiral decomposition and bosonization of both the explicit dimerization, Eq. \ref{eq:dim1} and the dimerized interaction, Eq. \ref{eq:dimstag}, it turns out they both give the same term in the low energy theory\cite{cgn07}
\bea
\ham_{\rm dim} &=& g_\Delta {\cal O}_{\rm dim} \nn
 {\cal O}_{\rm dim} &=& (4/\pi\alpha) 
  [ \cos(\sqrt{\pi}\Phi_c^+) \cos(\sqrt{\pi}\Phi_c^-) \cos(\sqrt{\pi}\Phi_s^-) 
\cos(\sqrt{\pi}\Phi_s^+)  \nn &+&  \sin(\sqrt{\pi}\Phi_c^+) \sin(\sqrt{\pi}\Phi_c^-) 
\sin(\sqrt{\pi}\Phi_s^-) \sin(\sqrt{\pi}\Phi_s^+)]. \label{eq:dimoperator}
\eea
where $g_\Delta=V/2n+O(1/n^2)$ is the sum of both the curvature and interaction terms.  This term must be added to the Kinetic part, the long range Coulomb part and the umklapp part to obtain the full Hamiltonian for zigzag nanotubes
\be
\ham = \ham_{\rm kin} + \ham_{\rm Coul} + \ham_{\rm umkl} + \ham_{\rm dim}\label{eq:ham-eff-total}
\ee

This Hamiltonian cannot be solved in general, so we consider the physical limit $U \gg V$.  The dimerization $\ham_{\rm dim}$ may then be considered a perturbation on the state created by the umklapp term.  As in the previous section, the large $U$ in the umklapp term locks the total charge field $\Phi_c^+=0$, and one can again perform the adiabatic approximation.  Noting that $\la \cos (\sqrt{\pi}\Phi_c^+) \ra \ne 0$ whereas $\la \sin (\sqrt{\pi}\Phi_c^+) \ra \ne 0$, one can consider only the first term of the dimerization operator in Eq. \ref{eq:dimoperator}, which after refermionization acquires the following low energy form
\be
{\cal O}_{\dim} \sim \cos(\sqrt\pi \Phi^- _c) \mu_0 \mu_1 \mu_2 \mu_3.\label{dimreferm}
\ee
In the spin liquid phase (see previous section), the operator can be further simplified by replacing the operators $\cos(\sqrt\pi \Phi_c^-)$ and $\mu_0$ by
their expectation values.  The resulting operator is:
\bea
\ham_\Delta &=& h_{\reff} ~ \mu_1 \mu_2 \mu_3, \nn
h_{\reff} &=& \frac{4g_\Delta}{(\pi \alpha)^2}
\langle \cos(\sqrt{\pi}\Phi_c^+)\rangle  \langle \cos(\sqrt{\pi}\Phi_c^-)\rangle
\langle \mu_0 \rangle. \label{heff}
\eea
All of the physics is now in the spin sector - in some sense, the charge sector of the Mott insulating ground state is {\em compatible} with the dimerization operator - both being happily minimized by the same state.  Thus the resulting theory is that of a dimerized spin ladder - for a cartoon account of such a system, see Appendix \ref{sec:snake}.

We see that the dimerization term competes with umklapp processes which support
the ground state with $ \la \sigma_i \ra \ne 0 $, $ \la \mu_i \ra = 0$.   The cartoon picture shows that as $g_\Delta$ is increased to a critical value, an $SU(2)_1$ critical point (i.e. the criticality of the Heisenberg spin {\em chain}) will ensue.  Such a claim can be substantiated by various theoretical techniques, such as a mapping to a non-linear sigma model\cite{mdss96}  or a mapping to a generalized Ashkin-Teller model.\cite{wn00}

The critical point may be estimated as the point where the mass generated by the dimerization term alone is equal to the triplet mass.  Although both terms originate from the nearest neighbour interaction $V$, they have different scaling dimensions, and a simple calculation\cite{cgn07} gives
\be
\frac{m_\Delta}{|m_t|} \propto \left(\frac{\bar{U}}{\bar{V}}\right)^{3/13} 
\left(\frac{n}{\bar{V}}\right)^{2/13}, 
\ee
where $\bar{U} = U \alpha/v$ and $\bar{V} = U \alpha/v$ are dimensionless coupling constants.  This shows that for sufficiently large radius $n$, the dimerization term will always win, and the ground state will be a dimerized phase, a phenomena completely unexpected from the $1/n^2$ scaling of the bare dimerization term.  Whether the quantum phase transition from the spin liquid to the dimerized phase occurs or not as a function of $n$ depends on various non-universal prefactors and is at present an open question.

\subsection{Long range Coulomb + single particle gap}

Another limit of the Hamiltonian Eq. \ref{eq:ham-eff-total} may be solved - when one neglects the umklapp processes and considers only the dimerization (either explicit of interaction induced) and the long range Coulomb terms.\cite{lt03}  It is not clear whether such a situation could be realized in nanotubes - but it is nevertheless an instructive theoretical exercise in the general properties of the Hamiltonian Eq. \ref{eq:ham-eff-total}.

The total Hamiltonian now is
\bea
\ham &=& \frac{v_\rho^+}{2} \int dx \,
\left[
\frac{1}{K} \left( \p_x \Theta_\rho^+ \right)^2 +
K \left( \p_x \Phi_\rho^+ \right)^2 \right]
+ \frac{v_F}{2} \sum_{a=(c^-,s^\pm)} \left[ \left( \p_x \Theta_a \right)^2
+ \left( \p_x \Phi_a \right)^2 \right] \nn
&+& (4g_\Delta/\pi\alpha) 
  [ \cos(\sqrt{\pi}\Phi_c^+) \cos(\sqrt{\pi}\Phi_c^-) \cos(\sqrt{\pi}\Phi_s^-) 
\cos(\sqrt{\pi}\Phi_s^+)  \nn && \;\;\;\;\;\;\;\; +  \sin(\sqrt{\pi}\Phi_c^+) \sin(\sqrt{\pi}\Phi_c^-) 
\sin(\sqrt{\pi}\Phi_s^-) \sin(\sqrt{\pi}\Phi_s^+)]
\eea
One still has the case that the total charge field $\Phi_c^+$ is much faster than the other fields, so the adiabatic approximation may be deployed.  However in this case, the field is not simply locked at $\Phi_c^+=0$ - both the cosines and sines in the dimerization term have the same scaling dimension so both must be treated equally.  It turns out that on integrating out the fast $\Phi_c^+$ mode, one is left with an $O(6)$ Gross-Neveau model\cite{lt03}, which is integrable\cite{zz79} and has a spectrum consisting of $3+6+3$ relativistic massive particles transforming as different representations of $O(6)$.  More details about this limit are beyond the scope of this review, we refer the interested reader to the original research article.\cite{lt03}

\section{The state of experiments}\label{sec:expt}

Transport experiments on nanotubes have provided some of the best experimental data of the Luttinger liquid state to date.  The experiments in 1999 of Bockrath et al\cite{bock99} clearly show power laws in conductivity as a function of either bias voltage or temperature, and all of the data collapses onto a universal scaling curve well described by Luttinger liquid theory.  Furthermore, as described in Section \ref{sec:Coulomb}, the Luttinger liquid parameter can be estimated from mesoscopic effects, and the difference in the exponent between tunneling contacts at the end of the nanotube or in the bulk of the nanotube can be clearly seen.  Although this experiment was actually performed on ropes of nanotubes, it was shown beforehand\cite{bock97} that transport in the ropes is dominated by a single metallic nanotube.

Experiments on isolated single wall carbon nanotubes are more difficult, although there have been several important studies.\cite{tans97,tans98,cobden98,liang02}  Although these studies focussed much more on mesoscopic effects such as Coulomb blockade, they clearly show the importance of electron-electron interactions.

More recently, a spin gap in carbon nanotubes has also been measured for the first time by means of an ingenious NMR technique.\cite{sing05,dgsk07}  In fact, even away from the case of half-filling discussed in this paper, interactions are expected to open a spin gap.\cite{eg98}  This gap would be much smaller than those predicted by the undoped analysis, as they come from marginally relevant backscattering and interband forward scattering terms (ignored in this paper), rather than the strongly relevant umklapp ones.  We finally comment that to date, neither the Mott insulating phase nor any of the unusual properties of the spin liquid phase have been seen experimentally.

\section{Conclusions and future directions}

\begin{figure}
\begin{center}
\includegraphics[width=3in]{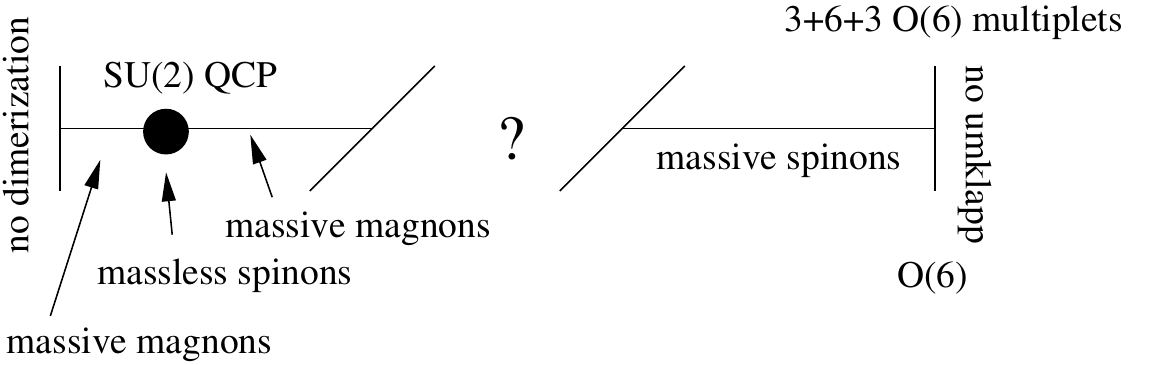}
\caption{The phase diagram between the known limits}\label{fig:phasediagram}
\end{center}
\end{figure}

We have presented a pedagogical overview of the theory of strong correlations in metallic single wall carbon nanotubes.  We particularly concentrated on the undoped case, where the combination of the unscreened Coulomb interaction, umklapp terms and a staggered interaction lead to a rich phase diagram - Fig. \ref{fig:phasediagram}, which is only partly understood.

It is well known that a single chain Mott insulator transforms to a band insulator when a dimerization term is switched on via not one but two seperate phase transitions,\cite{fgn99} one in the spin sector and one in the charge sector.  The phase diagram of Fig. \ref{fig:phasediagram} is reminiscent of this - for small dimerization there is one phase transition in the spin-sector at low dimerization, however we know at large dimerization (and no Mottness) that the charge is aranged differently.  Whether this is reached from a crossover or another quantum phase transition is an open theoretical question.  In fact, there are many open questions in the physics of dimerized ladder models - different sorts of dimerization may be added (whether relevant for nanotubes or not), and it may also be interesting to know how much of a difference screening the long range Coulomb interaction would make to this picture.

Returning to models directly relevant to nanotubes, it is known that the umklapp terms (leading to the spin liquid phase) are universal and should be roughly the same in all chiral nanotubes as well the as armchair and zigzag cases.  However, no study has been performed looking at the dimerization of interactions in the general chiral case.  One might expect chiral to be intermediate between the undimerized armchair case, and the maximally dimerized zigzag case, with the chiral angle being a plasuible  parameter to tune nanotubes to the $SU(2)_1$ quantum critical point.

On the experimental side - nanotubes have already provided some of the nicest experimental evidence for the Luttinger liquid state existing in nature.  However, the predicted Mott insulating phase has never been observed.  This may be because the nanotubes are naturally doped away from half filling, or possibly the Mott insulating state has been seen but the charge gap has been mistaken for one of the semiconducting nanotubes.  There is as yet no 'smoking gun' experiment to distinguish between these two possibilities on an individual nanotube.

In summary, we have presented an overview of the current (theoretical) understanding of strong correlation effects in carbon nanotubes.  With the current rate of improvement in experimental manipulation and observation of nanotubes, this is a ripe topic with plenty more unanswered questions.

\section*{Acknowledgements}

STC is supported by EPSRC grant no. GLGL RRAH 11382, and would like to thank S.Flatres for suggestions whilst preparing the manuscript.

\appendix{Bosonization}\label{sec:boz}

A full account of the bosonization technique is not possible, nor relevant for a review of this nature.  However, in order to make this article self-contained and to define our notations, we present a brief description here.  There are a number of good references\cite{boz1,boz2,boz3} for the reader who wishes to delve deeper into this subject.

The basic idea behind bosonization is that in a model with a linear spectrum, excitations consist of particles and holes moving with the same velocity which are therefore coherent.  The entire spectrum can thus be reformulated in terms of these bosonic particle-hole excitations, i.e. the local density operators, or in other words, the {\em plasmons}.  This is a very useful representation, as the density-density interaction remains quadratic, thus allowing the problem to be solved.

To be more technical, consider the fermionic Hamiltonian
\be
\ham_F = -iv_F (R^\dagger \p_x R - L^\dagger \p_x L), \label{eq:hf}
\ee
which has the single particle propagators (in imaginary time)
\be
\la \mbox{T}_\tau R (x,\tau) R^\dagger (0,0) \ra = \frac{1}{2\pi} \frac{1}{x+i\tau}, \;\;\;
\la \mbox{T}_\tau L (x,\tau) L^\dagger (0,0) \ra = \frac{1}{2\pi} \frac{1}{x-i\tau}.
\ee
Now, consider the bosonic Gaussian Hamiltonian
\be
\ham_B = \frac{v_F}{2} \left[ \Pi^2(x) + (\p_x \Phi (x) )^2 \right],\label{eq:hb}
\ee
which has the single particle propagator
\be
\la \mbox{T}_\tau \Phi(x,\tau) \Phi(0,0) \ra = \frac{1}{4\pi} \ln \left( \frac{R^2}{x^2+\tau^2} \right)
= \frac{1}{4\pi} \left(\frac{R}{x+i\tau}\right) + \frac{1}{4\pi}\left(\frac{R}{x-i\tau}\right), \label{eq:bos_green}
\ee
where $R$ is the long distance (low energy) cutoff (i.e. the system size).  From this expression, we see that at the level of correlation functions, one can decompose the field
\be
\Phi(x,\tau) = \phi_R (x+i\tau) + \phi_L(x-i\tau),\label{eq:conf}
\ee
i.e. into a 'right moving' part that depends only on the combination $x+i\tau$ and a 'left moving' part depending only on $x-i\tau$.  A short calculation\cite{boz1} shows that
\be
 \la \mbox{T}_\tau e^{i\beta \Phi(x,\tau)} e^{-i\beta \Phi(0,0)} \ra 
=
e^{ \beta^2 \la  \Phi(x,\tau) \Phi(0,0) \ra } e^{ -\beta^2 \la \Phi(0,0) \Phi(0,0) \ra}
= \left( \frac{a_0^2}{x^2 +\tau^2}\right)^{\beta^2/4\pi},\label{eq:exponentcorr}
\ee
where $a_0$ is a short distance (high energy) cutoff (i.e. the lattice spacing).  With this expression and the decomposition (\ref{eq:conf}), we see that if we define the correspondence
\be
R(x) = \frac{1}{\sqrt{2\pi a_0}} e^{ i\sqrt{4\pi} \phi_R (x) }, \;\;\;\;\;
L(x) = \frac{1}{\sqrt{2\pi a_0}} e^{- i\sqrt{4\pi} \phi_L (x) }, \label{eq:operators}
\ee
then the correlation functions between the fermionic model and the bosonic exponents are identical.  In fact, it turn out\cite{boz1}  that the entire spectrum of models \ref{eq:hf} and \ref{eq:hb} is the same, so the operator mapping \ref{eq:operators} can be considered an exact equivalence between the two models, including when further perturbations such as interactions are added to the Hamiltonians.  The final ingredient to be added is the correct way to calculate the density
\be
\rho(x) = R^\dagger (x) R(x) + L^\dagger(x) L(x) = \frac{1}{\sqrt{\pi}} \p_x \Phi(x).
\ee
It is often useful to introduce the field
\be
\Theta(x) = -\phi_R(x) + \phi_L(x) \label{eq:thetadef}
\ee
which satisfies $\p_x \Theta(x) = \Pi(x)$.  The fields $\Phi(x)$ and $\Theta(x)$ satisfy the non-local commutation relation
\be
\left[ \Phi(x), \Theta(y) \right] = i\theta(x-y).
\ee
where $\theta(x-y)$ is the Heaviside function.  Expressing the Hamiltonian in terms of $\Phi(x)$ and $\Theta(x)$
\be
\ham_B = \frac{v_F}{2} \left[ (\p_x \Theta(x) )^2 + (\p_x \Phi (x) )^2 \right],
\ee
it is easy to see that $\Theta(x)$ is the dual field to $\Phi(x)$, and therefore, the propagator $\la \mbox{T}_\tau \Theta(x,\tau) \Theta(0,0) \ra$ is also given by Eq. \ref{eq:bos_green} identical to the $\Phi$ case.

We now consider perturbing the Hamiltonian \ref{eq:hf} or \ref{eq:hb} by a density-density interaction
\be
\ham' = g\rho(x) \rho(x+a).
\ee
In the bosonic case, the Hamiltonian remains quadratic
\bea
\ham_B + \ham' &=& \frac{v_F}{2} \left[ (\p_x \Theta(x) )^2 + \left( 1+ \frac{g}{\pi v_F} \right) (\p_x \Phi (x) )^2 \right] \nn
&=& \frac{v'}{2} \left[ K (\p_x \Theta(x) )^2 +  \frac{1}{K} (\p_x \Phi (x) )^2 \right],
\eea
where $K =1/ \sqrt{1+g/\pi v_F}$ is known as the {\em Luttinger liquid parameter} and $v'=v_F/K$ is the renormalized velocity.  This equation can therefore be brought back to canonical form by the shift of variables $\Phi = \sqrt{K} \Phi'$, $\Theta = \Theta'/\sqrt{K}$.  One can then calculate fermionic correlation functions in the interacting model
\bea
&& \la \mbox{T}_\tau R(x,\tau) R^\dagger (0,0) \ra 
\sim
 \la \mbox{T}_\tau e^{i\sqrt{\pi}[\Phi(x,\tau)-\Theta(x,\tau)]} e^{-i\sqrt{\pi} [\Phi(0,0) - \Theta(0,0)]} \ra  \nn
&=& \la \mbox{T}_\tau e^{i\sqrt{\pi K} \Phi(x,\tau) - i\sqrt{\pi/ K} \Theta(x,\tau)} e^{-i\sqrt{\pi K} \Phi(0,0) + i\sqrt{\pi/ K} \Theta(0,0)}  \ra   \nn
&=&  \la \mbox{T}_\tau e^{ i [ \sqrt{\pi K} + \sqrt{\pi/K} ] \phi_R (x,\tau) }
e^{- i [ \sqrt{\pi K} + \sqrt{\pi/K} ] \phi_R (0,0) } \ra
\nn && \times 
\la \mbox{T}_\tau e^{ i [ \sqrt{\pi K} - \sqrt{\pi/K} ] \phi_L (x,\tau) }e^{ i [ \sqrt{\pi K} - \sqrt{\pi/K} ] \phi_L (0,0) } \ra \nn
&=&
\left(\frac{a_0}{x+i\tau}\right)^{\frac{1}{4}(\sqrt{K}+\sqrt{1/K})^2}
\left(\frac{a_0}{x-i\tau}\right)^{\frac{1}{4}(\sqrt{K}-\sqrt{1/K})^2} \nn
&=&  \frac{a_0}{x+i\tau} \left(\frac{a_0^2}{x^2+\tau^2}\right)^{(K+1/K-2)/4}.
\eea
The Green's function no longer shows a quasi-particle pole at $x=-i\tau$ - in a Luttinger liquid, all single particle excitations are incoherent.

\subappendix{More than one species of Fermion}

The previous section describes the bosonization of a single species of Fermion, however in the low-energy effective theory of nanotubes, there are 4 species - 2 spins and 2 bands.  The non-interacting Hamiltonian used as a starting point is
\be
\ham_F =  \sum_{\mu\sigma} -iv_F (R^\dagger_{\mu\sigma} \p_x R_{\mu\sigma} - L^\dagger_{\mu\sigma} \p_x L_{\mu\sigma}),
\ee
where $\mu=\pm$ is the band index and $\sigma=\uparrow,\downarrow$ is the spin index.  Bosonization takes place individually for each $\mu\sigma$ combination exactly analagous to the previous section for the four fields $\Phi_{\mu\sigma}$
\bea
R_{\mu\sigma} (x) = \frac{\kappa_{\mu\sigma}}{\sqrt{2\pi a_0}} \exp \left\{ -i\sqrt{\pi}\left[ \Phi_{\mu\sigma}{x} - \Theta_{\mu\sigma}(x) \right] \right\}, \\ \nonumber
L_{\mu\sigma} (x) = \frac{\kappa_{\mu\sigma}}{\sqrt{2\pi a_0}} \exp \left\{ -i\sqrt{\pi}\left[ \Phi_{\mu\sigma}{x} +\Theta_{\mu\sigma}(x) \right] \right\},
\eea
where the $\kappa_{\mu\sigma}$ are anti-commuting variables (known as Klein factors) which ensure the anti-commutation between the different species of Fermions.  In the case of nanotubes, these variables have no dynamics, so for most purposes can be safely ignored - their only role is to ensure the correct sign of backscattering terms and operators representing possible order paramters.  We will say no more about them in this review, and refer the reader to the original research papers for more details about their role in the theory of nanotubes.

It turns out that instead of the fields $\Phi_{\mu\sigma}$ which represent density fluctuations in the four different fermionic species, the linear combinations
\bea
\Phi_c^\pm = \frac{1}{2} \left( \Phi_{+\uparrow}+ \Phi_{+\downarrow} \pm \Phi_{-\uparrow} \pm \Phi_{-\downarrow} \right), \\ \nonumber
\Phi_s^\pm = \frac{1}{2} \left( \Phi_{+\uparrow}- \Phi_{+\downarrow} \pm \Phi_{-\uparrow} \mp \Phi_{-\downarrow} \right).\label{eq:linear}
\eea
which decribe fluctuations in total charge $(c^+)$, relative charge $(c^-)$, total spin $(s^+)$ and relative spin $(s^-)$ are a much better basis when interactions are added.  In this representation, the non-interacting Hamiltonian remains the sum of four independent Gaussian models
\be
\ham_{kin} = \frac{v_F}{2} \sum_{a=(c^\pm,s^\pm)} \left[ \left( \p_x \Theta_a \right)^2
+ \left( \p_x \Phi_a \right)^2 \right]. 
\ee
As an intuitive example, let's see what the single-particle propagator looks like when we add the long-range Coulomb interaction - which affects only the total charge channel (Eq. \ref{eq:Hcoul} in the main text):
\be
\la \mbox{T}_\tau R_{+,\uparrow}(x,\tau) R_{+,\uparrow}^\dagger (0,0) \ra 
\sim \prod_a \la \mbox{T}_\tau e^{i\sqrt{\pi/4}[\Phi_a(x,\tau)-\Theta_a(x,\tau)]} e^{-i\sqrt{\pi/4} [\Phi_a(0,0) - \Theta_a(0,0)]} \ra,
\ee
where $a=c^+,c^-,s^+,s^-$.  The results of the previous section can then be applied to each of the four sectors individually as each of the sectors commutes with the other ones.  The only difference is the coefficient $\beta$ in the exponent $e^{i\beta\Phi}$ is now $\sqrt{\pi/4}$ rather than $\sqrt{\pi}$ because of the normalization of the four fields.  The result is
\be
\la \mbox{T}_\tau R_{+,\uparrow}(x,\tau) R_{+,\uparrow}^\dagger (0,0) \ra 
\sim \left(\frac{a_0}{v_\rho^2 x^2 + \tau^2}\right)^{(K+1/K-2)/16}
\left(\frac{a_0}{ v_\rho x + i\tau}\right)^{1/4} \left(\frac{a_0}{v_F x + i\tau}\right)^{3/4},\label{eq:fullcorr}
\ee
the first two terms arising from the $c^+$ sector, and the remaining term being the product of the other three sectors of the theory.  Similar results arise for each of the other possible propagators.

\subappendix{Tunneling Density of States}\label{sec:tdos}

The tunneling density of states is calculated as the Fourier transform of the imaginary part of the retarded Green function
\be
\nu (\omega) = \frac{-1}{\pi} \int e^{i\omega t} \, \Im \, G^R(x=0,t) dt.
\ee
The retarded Green function is obtained from the Matsubara one (\ref{eq:fullcorr}) by the analytic continuation $\tau=it+\delta$, so it is seen that
\be
G^R(x=0,t) \sim t^{-(K+1/K-2)/8+1}
\ee
is a power law.  Now,
\be
\int e^{i\omega t} t^{-\alpha} dt \sim \omega^{\alpha-1},
\ee
so the density of states
\be
\nu(\omega) \sim \omega^{(K+1/K-2)/8}
\ee
as stated in the main text.

So far, everything we have described is for an infinite homogenous system - which is a good approximation so long as we stay in the middle of the nanotube.  If we probe the edge, we must worry about the presence of the boundary.  Fortunately this is simple within bosonization: no particles can leave, so at the right edge, a right mover must be reflected as a left mover, i.e. $R^\dagger(x_{max})=L(x_{max})$.  From Eq. \ref{eq:operators}, we see that this means $\phi_R(x_{max})=\phi_L(x_{max})$, and from Eq. \ref{eq:thetadef}, this implies that $\Theta=0$ at the boundary.  Therefore, at the boundary $x_m$, Eq. \ref{eq:fullcorr} is modified to exclude correlations of $\Theta$:
\bea
\la \mbox{T}_\tau R_{+,\uparrow}(x_m,\tau) R_{+,\uparrow}^\dagger (x_m,0) \ra 
&=&  \prod_a \la \mbox{T}_\tau e^{i\sqrt{\pi/4}[\Phi_a(x_m,\tau)]} e^{-i\sqrt{\pi/4} \Phi_a(x_m,0) } \ra
\nn
&\sim& (i\tau)^{(1/K-1)/4+1}.
\eea
which gives the boundary density of states
\be
\nu(\omega) \sim \omega^{(1/K-1)/4}.
\ee

\subappendix{The sine-Gordon model}\label{sec:sG}

In the previous sections we showed that density-density interactions in the single species case left the bosonic Hamiltonian quadratic, and therefore gave rise to a Luttinger liquid, characterized by power law decays of correlation functions.  When either interband  backscattering
 (e.g. $R^\dagger_{+,\sigma} R_{-,\sigma} L^\dagger_{-,\sigma} R_{+,\sigma}$),
 spin-flip backscattering
  (e.g. $R^\dagger_{\mu,\uparrow} R_{\mu,\downarrow} L^\dagger_{\mu,\downarrow} R_{\mu,\uparrow}$),
 or umklapp terms
(e.g. $R^\dagger_{\mu,\uparrow} L_{\mu,\downarrow} R^\dagger_{\mu,\downarrow} L_{\mu,\uparrow}$)
  are taken into account, this is no longer true, and cosine terms are generated in the Hamiltonian.
  
While in the main text, we study the full Hamiltonian for carbon nanotubes which include coupling between the channels, here we discuss the basic unit which appears in a number of places: the canonical sine-Gordon model in a single channel
\be
\ham _{sG}=  \frac{v_F}{2} \left[ \Pi^2(x) + (\p_x \Phi (x) )^2 \right] - \frac{g}{(\pi a_0)^2} \cos (\beta \Phi(x) ).\label{eq:sG}
\ee
The cosine term would like to lock the field $\Phi$ at the minimum of the potential $\Phi(x)=0$ is $g>0$ and $\Phi(x)=\pi/\beta$ if $g<0$, however the momentum term $\Pi^2(x)$ is minimized when $\Phi(x)$ has large temporal fluctuations.  The relevant importance of each of these terms is controlled by the parameter $\beta$.  The scaling dimension of the operator $\cos (\beta \Phi(x))$ can be read off from Eq. \ref{eq:exponentcorr}: $d=\beta^2/4\pi$.  When the scaling dimension is greater than $2$, correlation functions of this operator do not diverge (at low energy) when integrated over position and time.  A scaling dimension of less than $2$ indicates a divergence will occur when perturbation theory is applied - and so the operator may substantially change the ground state.

The sine-Gordon model is integrable, and is one of the most studied models in physics.\cite{sgrefs}  Here, we will simply give a summary of the main results without derivation.  If $\beta^2>8\pi$, then the momentum term wins, and the cosine perturbation fails to lock the field $\Phi(x)$ - in technical terms we say it is an irrelevant operator.  In this case, we can forget about it's presence and deal only with the Gaussian model.

If however, $\beta^2<8\pi$, then the cosine term wins, and the field $\Phi(x)$ is locked, giving a non-zero expectation value to $\la \cos(\beta/2 \Phi(x)) \ra \ne 0$ if $g>0$ and $\la \sin(\beta/2 \Phi(x)) \ra \ne 0$ if $g<0$.  There is a spectral gap to excitations, and the excitations are given by solitons in the field $\Phi(x)$, interpolating between two adjacent minima of the cosine potential.  These solitons can be visualized as in the cartoon Appendix. \ref{sec:soliton}.  An anti-soliton is then the mirror image of a soliton.

Furthermore, there is a residual interaction between solitons and anti-solitons - which is repulsive if $4\pi<\beta^2<8\pi$ and attractive if $\beta^2<4\pi$.  In the attractive case, the solitons and anti-solitons can form bound states (with an energy gap less than the individual solitons), which are once more similar to the density waves of the original Gaussian model - this is what happens in the total charge channel on the half-filled nanotubes (see main text) .  At exactly $\beta^2=4\pi$, the solitons are free and may once more be re-expressed in term of fermions.

\subappendix{Refermionization and Majorana fermions}\label{sec:refermionization}

Refermionization is due to the equivalence of a model of massive Dirac fermions to the $\beta^2 = 4\pi$ sine-Gordon model, demonstrated  by taking the Hamiltonian
\be
\ham_{sG-F} = -iv_F (R^\dagger \p_x R - L^\dagger \p_x L) + i g (R^\dagger L -L^\dagger R)
\ee
and bosonizing according to the rules \ref{eq:operators}.  The result is the sine-Gordon Hamiltonian \ref{eq:sG} (the cosine is obtained rather than a sine from the correct treatment of the product of the exponents of two non-commuting operators, $\Phi$ and $\Theta$)\cite{boz3}.  In this case, the bosonization can work backwards and is known as refermionization. The Fermions represent the non-interacting solitons of the sine-Gordon model.  It must however be remembered that the sine-Gordon model usually appears in one of the sectors of the theory invoked by the linear combination \ref{eq:linear}, so are not simply related to the original Fermions $R_{\mu\sigma}$ of the model (in fact the relationship is in general not even local).

When treating the spin sector of nanotubes, it is convenient to re-express the Dirac Fermions in terms of Majorana (real) fermions:
\bea
\chi^1_R = \frac{R^\dagger +R}{2}, &\;\;\;\; & \chi^2_R = \frac{R^\dagger-R}{2i}, \nn
\chi^1_L = \frac{L^\dagger +L}{2}, & & \chi^2_L = \frac{L^\dagger-L}{2i}.
\eea

The Hamiltonian of the Majorana fermions is then the kinetic part:
\be
\ham_{M} = -iv_F \sum_{a=1}^2  \left ( \chi_R^a \p_x \chi_R^a - \chi_L^a \p_x \chi_L^a \right),
\ee
and the possible mass terms are
\bea
 \chi_R^1 \chi_L^1 + \chi_R^2 \chi_L^2 &=& i (R^\dagger L - L^\dagger R )= \frac{1}{(\pi a_0)^2}\cos(\sqrt{4\pi} \Phi ) \nn
  \chi_R^1 \chi_L^1 - \chi_R^2 \chi_L^2 &=& R^\dagger L^\dagger + L R = \frac{1}{(\pi a_0)^2} \cos(\sqrt{4\pi} \Theta ) \nn.
\eea
In the main text, the refermionization of the spin sector in terms of Majorana fermions comes about by carrying out this procedure for the two sectors $\Phi_s^+$ and $\Phi_s^-$, thus arriving at 4 Majorana fermions.

The utility of this representation comes about from the equivalence between a single species of Majorana fermion and the transverse field Ising model
\be
\ham_I = -J \sum_i  \sigma^z_i \sigma^z_{i+1} + h \sum_i \sigma^x_i
\ee
after carrying out a Jordan-Wigner transformation.\cite{boz1,snt96}  The mass of the Majorana comes out to be $m=|J|-|h|$, with the $Z_2$ quantum critical point at $|J|=|h|$.  If $|J|>|h|$, then this model is in it's ordered phase $\la \sigma^z \ra \ne 0$, whereas in the opposite case $|h|>|J|$, the so called {\em disorder parameter}
$
\mu_i^z= \prod_{j<i} \sigma^x_j
$
is non-zero $\la \mu^z \ra \ne 0$.  By bosonizing two copies of the Ising model, one can relate $\sigma$ and $\mu$ to the bosonic fields $\Phi$ and $\Theta$.\cite{boz1,snt96}  For details of this, see the literature - the correspodence we use is shown in the main text, and allows the strong coupling phases of carbon nanotubes to be understood.

\appendix{Cartoon pictures of the physics of Mott insulators, Spin chains and Spin ladders}\label{sec:cartoon}

Here, we provide a strong coupling 'cartoon' picture of the physics of the correlated states of carbon nanotubes.  These are provided both as a way to understand the one-dimenionsal physics involved without delving into the details of bosonization, and as a way of visualizing the underlying field theory.  We stress that these are cartoons and are no substitute for a proper calculation, however we feel that they give a very intuitive window into the physics of one dimensional systems, which is sometimes lost in the technicalities of field theory.

\subappendix{One dimensional interacting electrons}\label{sec:plasmons}
First, consider electrons confined to move in only one dimension, i,e. on a line. Even with arbitrarily small interactions, the electrons cannot go {\em through} each other, therefore there are no (coherent) single particle electron like excitations, and all excitations are plasmons.
\begin{center}
\includegraphics[width=3in]{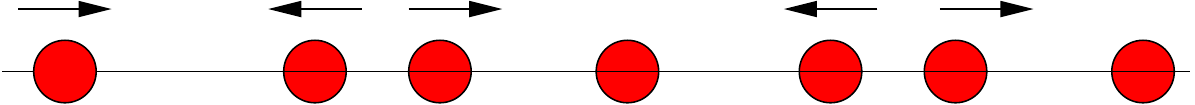}
\end{center}
This is a basic picture of the Luttinger liquid state.

\subappendix{Half filling in a lattice - the Mott insulator}\label{sec:mott}
Now consider a one-dimensional lattice model - the Hubbard model with an energy cost $U$ for two particles to be on the same site, and a nearest neighbor hopping integral $t$.  At half filling, the electrons want to avoid the energy cost $U$, so have exactly one electron per site - the Mott insulator.  It turns out that in contrast to higher dimensions, a one dimensional system will form a Mott insulator at half filling for arbitrarily small on-site interaction $U$.
 \begin{center}
\includegraphics[width=3in]{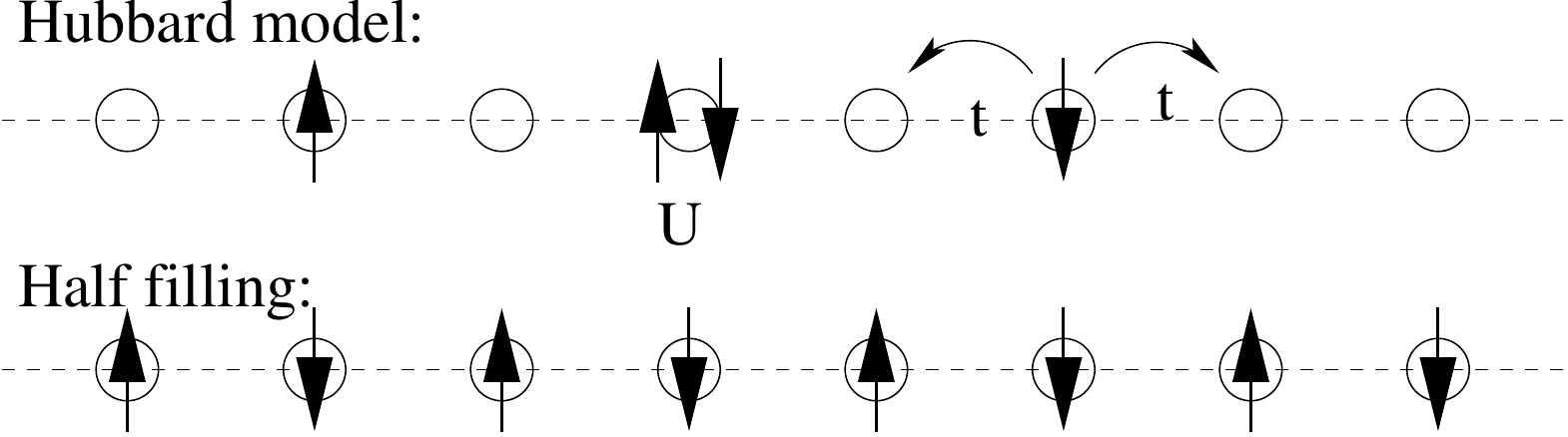}
\end{center}
There is a gap of order $U$ to any charged excitation, but there is a residual antiferromagnet exchange interaction of order $t^2/U$ from virtual hopping terms.  All of the interesting low-energy physics is in the spin sector - so we need to consider the physics of one dimensional spin chains. 

\subappendix{Spin Chain - deconfined solitons}\label{sec:soliton}
Consider making an excitation in an antiferromagnetic spin chain by flipping the spin in the center - this changes the spin of the system by $1$ and creates two {\em unsatisfied} bonds - as shown in figure (a). These unsatisfied bonds however can move independently, splitting into two {\em solitonic} excitations known as {\em spinons}, each carrying a spin $1/2$ - figure (b).
\begin{center}
\includegraphics[width=4in]{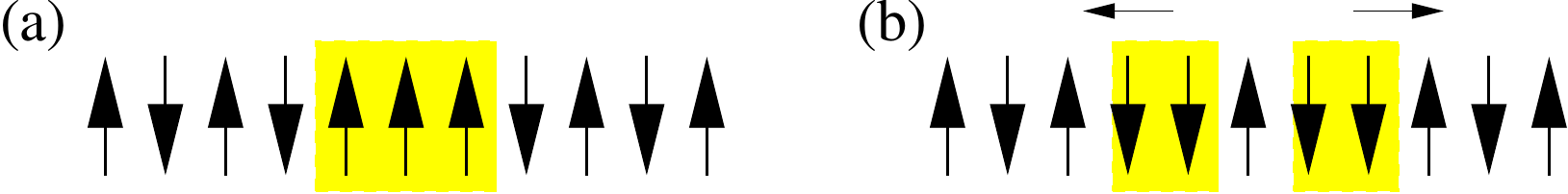}
\end{center}
We must stress here again that this is a cartoon picture - in fact, the Heisenberg spin chain has only quasi-long-range order (i.e. power law correlations), even at zero temperature, and so never actually forms the Neel state pictured.  Nevertheless, the basic physics is represented by the cartoon is correct.

\subappendix{Spin Ladder - confinement of solitons}\label{sec:confinement}
At this stage, we must remember that the low-energy model relevant for nanotubes is actually a ladder model, and not a single chain.  While this makes very little difference in the charge sector - at half filling the ladder is still a Mott insulator, the physics of the spin ladder turns out to be very different from the spin chain.
\begin{center}
\includegraphics[width=2in]{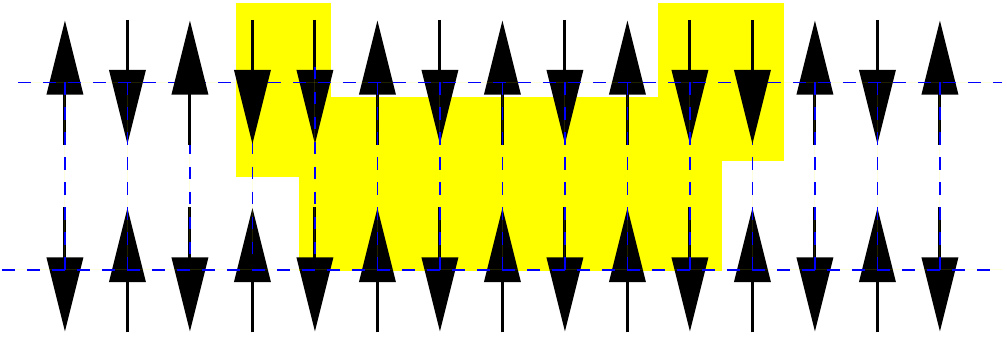}
\end{center}
Suppose two solitons form on one of the legs of the ladder and start to move apart.  Then all of the bonds from the first leg to the second between the two solitons are unsatisfied - thus as the solitons move further apart, the energy increases linearly creating a confining potential (actually, very similar in style to the confining potential between two quarks in quantum chromodynamics).  The excitations can then roughly be split into two classes: either the two solitons pictured must move together, which is a very rough cartoon of the {\em triplet} excitations, or a soliton on one chain moves coherently with an antisoliton on the other chain - which is very roughly the {\em singlet} excitation.  It turns out that the confinement energy leads to both of these classes of excitations being gapped - although again one must remember that in reality, there is only quasi-long-range order and the Neel state is not truly formed.

\subappendix{Dimerized spin ladder - quantum criticality}\label{sec:snake}
Finally, we consider the dimerized spin ladder important for the zig-zag nanotubes.  In the absence of dimerization, the model is that of a uniform spin ladder, as pictured in section \ref{sec:confinement}, which has gapped triplet and singlet excitations (or in other words, excitations consisting of confined spinons).  However, now turn on the dimerization and consider tuning it very carefully as shown in the figure.
\begin{center}
\includegraphics[width=4in]{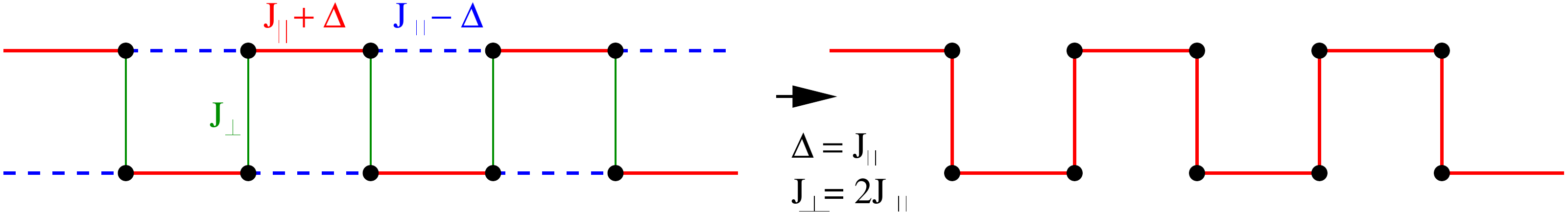}
\end{center}
At the critical point, the ladder becomes equivalent to a single chain - \ref{sec:soliton} - which has gapless excitations consisting of deconfined solitons.  This is the so called $SU(2)_1$ quantum critical point expected in zig-zag nanotubes.  Note that a full calculation shows that the critical point is in fact not just a point as in the cartoon, but an entire line $\Delta_c(J_\|,J_\perp)$.

\end{document}